\documentclass[epsfig,12pt]{article}
\usepackage{rotating}
\usepackage{amssymb,amsmath}
\usepackage{color}
\usepackage{epsfig}
\begin{document}

\begin{center}
\begin{large}
{\bf Prospects for Charged Particle Astronomy}\\
\end{large}
\vspace{.75cm}
{\bf Paul Sommers\\}
{\it Center for Particle Astrophysics and\\
Penn State Department of Physics\\
104 Davey Lab\\
University Park, PA 16802}
\end{center}

\vspace{1cm}
\begin{abstract}
The likelihood of detecting individual discrete sources of cosmic rays
depends on the mean separation between sources.  The analysis here
derives the minimum separation that makes it likely that the closest
source is detectable.  For super-GZK energies, detection  is signal
limited and magnetic fields should not matter.  For sub-GZK energies,
detection is background limited, and intergalactic magnetic fields
enter the analysis through one adjustable parameter.  Both super-GZK
and sub-GZK results are presented for four different types of sources:
steady isotropic sources, steady jet sources, isotropic bursts, and
jet bursts.
\end{abstract}

\begin{center}
Keywords: cosmic ray; anisotropy; propagation; discrete source\\
\end{center}

\section{Introduction}
The rapid growth of the Pierre Auger Cosmic Ray Observatory's
data set has prompted speculation that one or more
discrete sources should soon be detected at ultra-high cosmic ray
energies.  Just as nearby stars are easily detected against the night
sky, any nearby cosmic ray source should manifest itself as a cluster
of cosmic ray arrival directions in the sky.  At sufficiently high
energies, magnetic deflections of charged particles from nearby
sources should be small, so tight clustering of those arrival
directions can be expected.  At the highest energies, all of the
observed cosmic rays should be coming from nearby sources because pion
photoproduction (the GZK effect) eliminates high energy protons
from distant sources.

It is presently not possible to predict with certainty whether or not
new observatories (e.g. Auger \cite{Auger}, Telescope Array \cite{TA}) will identify and study
individual discrete cosmic ray sources.  The nature of the primary
particles is still unknown and so is the mechanism that endows them
with extremely high energy.  Even if it is assumed that the highest
energy cosmic rays are protons produced by discrete extragalactic
sources, those sources might be distributed densely like galaxies or
widely separated like the most powerful quasars.  The mean separation
between sources is crucial in evaluating the probability that the
nearest source is detectable.

Assumptions about intergalactic magnetic field properties are also
critical in evaluating this probability.  Proton trajectories are bent
by magnetic fields, so the flux of particles from a discrete source is
spread over a solid angle of the sky.  The amount of blurring due to
intervening extragalactic fields depends on the distance to the
source, on the typical magnetic field strength in intergalactic space,
and on the effective coherence length for magnetic field directions.
Properties of intergalactic magnetic fields are highly uncertain
\cite{Magfields,Dolag,Sigl}.  The better-known magnetic fields {\it
within} the Galaxy will blur any extragalactic source somewhat, but at
the highest energies the blurring of an extragalactic proton source
due to the Galaxy's fields is less than the detector's point spread
function (cf. Appendix B).

Prospects for charged particle astronomy may be greatest above the GZK
energy threshold for pion photoproduction \cite{GZK}.  Protons that
are now detected well above the GZK energy threshold must have been
produced within the last few hundred million years.  Their sources are
necessarily close enough, and their magnetic rigidities are great
enough, to expect that their arrival directions must point close to
the celestial positions of their sources.  Moreover, the GZK effect
eliminates the isotropic background which, at energies below the GZK
threshold, has been accumulating for roughly 5 billion years.

It is important to note that the favorable prospects for super-GZK
charged particle astronomy stem from two independent considerations.
It is a coincidence that magnetic blurring by galactic magnetic fields
and by plausible extragalactic fields becomes unimportant for protons
approximately at the threshold energy where the GZK effect erases the
isotropic background.  Magnetic fields have nothing to do with pion
photoproduction.  If the GZK energy threshold were lower, charged
particle astronomy would be favorable above the threshold even though
intervening magnetic fields might spread any one source over a large
solid angle on the sky, simply because the background would be gone.
Moreover, if magnetic fields were negligible below the GZK
threshold, point source excesses would be detectable despite
the isotropic background.  With the two effects taking hold at
approximately the same energy, it is assured that super-GZK astronomy
is signal-limited, whereas sub-GZK charged particle astronomy is
background-limited.

Neither the GZK erasure of isotropic background nor the defeat of
magnetic blurring by high rigidity has a sharp onset with energy.
Instead of designating any particular energy as the dividing line
between sub-GZK and super-GZK particle astronomy, this paper will rely
on the characterization of super-GZK as being signal-limited and
sub-GZK as background limited.  Signal-limited analysis will be called
super-GZK, whereas background-limited analysis will be called
sub-GZK.

The basic model to be used throughout most of this paper assumes that
ultra-high energy cosmic rays are produced by a set of sources with
mean separation $R$ and that all of them have the same luminosity $Q$
(cosmic rays per unit time).  The density of cosmic rays in the
universe today results from that emission rate density over an
accumulation time $T$ (which is limited by electron pair production
below the GZK threshold energy and by pion photoproduction above the
threshold).  The individual source luminosity $Q$ can be calculated
from $R$, $T$, and the observed cosmic ray intensity $I$.  For any
assumed distance to the nearest such source, the expected flux is then
determined.  Its detectability at that distance depends simply on the
cumulative detector exposure and, in the case of sub-GZK astronomy,
how much solid angle is spanned by the magnetically blurred flux.

It is easy to understand that the nearest source is more likely to be
detectable if the sources are widely separated.  Today's observed
cosmic ray density is $Q\times T/R^3$ (the density of sources being
$1/R^3$), so each source must have luminosity $Q\propto R^3$ in this
model.  A larger mean separation $R$ requires a greater luminosity $Q$
per source to account for the observed cosmic ray density.  The flux
from a single source at distance $r$ will be $Q/4\pi r^2 \propto
R^3/r^2$.  Since the nearest source is likely to be at distance $r\sim
R$, the expected flux from the nearest source grows linearly with $R$.
In the super-GZK (signal-limited) regime, this suffices to show that
sources are more detectable if the mean separation of sources $R$ is
large.  (This result pertains only when R is less than the GZK
attenuation length.)  For the sub-GZK case, one must worry about
magnetic fields enlarging the solid angle over which the signal is
seen by the detector.  If the magnetic smearing is stochastic, that
solid angle grows linearly with $r$ (and hence $R$), but the noise
(fluctuation of the background within that solid angle) grows only as
its square root.  The signal-to-noise (detectability criterion) is
therefore enhanced by large values of $R$ also in the sub-GZK regime.

The analysis proceeds to find the minimum separation $R$ that is
needed in order for there to be greater than a 50-50 chance that the
nearest source will be detectable.  The answer depends especially on
the assumed cosmic ray accumulation time $T$ for particles above an
adopted energy cut and on the number of arrival directions in the
detector's data set that are above that energy cut.  In the
background-limited case, the answer also depends on a parameter that
characterizes small-angle magnetic scattering in intergalactic space.
For any set of assumed values for these parameters, one obtains the
minimum mean separation of sources for which the detection of discrete
sources is likely.

The basic model assumes that the different sources of ultra-high
energy cosmic rays are identical.  They persist in time and emit
cosmic rays isotropically.  The model can be modified to study sources
that emit in collimated jets, or transient sources that emit their
cosmic rays isotropically in a burst, or bursts with jets.  Part of
the analysis can be performed with an arbitrary luminosity function
rather than assuming that all sources have the same luniosity.

No cosmic ray observatory has so far achieved full-sky
coverage.  Partial sky coverage is an obvious handicap in detecting
discrete sources, for the detector may not get exposed to the
brightest source.  In addition, the analysis becomes more complicated
without full-sky exposure.  Section 10 indicates the necessary
modifications for realistic calculations with partial sky coverage.  
The simplifying assumption in the bulk of this paper is that the
detector has the same exposure to sources anywhere in a fraction of
the sky denoted $f_D$.  For a detector at one mid-latitude site,
the effective fraction $f_D$ is approximately $1/2$ (cf. section 10).

\section{Notation}

\noindent\hangindent=25pt {\bf $E$}\hspace{.5cm} Cosmic ray energy,
measured in $EeV$ (1 $EeV$ = $10^{18} eV$).  In this paper, the
energy $E$ usually denotes the minimum energy used in the search for a
discrete source.  Due to the steep cosmic ray energy spectrum, most
of the included cosmic rays are not much above that minimum energy.

\noindent\hangindent=25pt {\bf $Q$}\hspace{.5cm} The luminosity of a
cosmic ray source (number of cosmic rays per unit time).

\noindent\hangindent=25pt {\bf $n$}\hspace{.5cm} The spatial density of
cosmic ray sources (number of sources per unit volume).  The meaning
of $n$ is somewhat different in the cases of bursting sources.  (See
the description of $\eta$ in this section.)

\noindent\hangindent=25pt {\bf $R$}\hspace{.5cm} The mean separation of
cosmic ray sources ($n=1/R^3$).  $R$ is measured in $Mpc$ ($3.26\times
10^6\ light\ years$).  

\noindent\hangindent=25pt {\bf $r$}\hspace{.5cm} The distance to a source.

\noindent\hangindent=25pt {\bf $r_0$}\hspace{.5cm} The distance to the
			  {\it nearest} source.

\noindent\hangindent=25pt {\bf $c$}\hspace{.5cm} The speed of light.

\noindent\hangindent=25pt {\bf $T$}\hspace{.5cm} The effective time
(measured in $years$) over which the cosmic rays have accumulated.
This depends on energy.  For energies below the GZK threshold that are
considered here, $cT$ is limited by $e^{\pm}$ pair production to
roughly $1500\ Mpc$.  Well above the GZK threshold, $cT$ is roughly
$30\ Mpc.$ The energy dependence of $T$ through the threshold energy
region is governed largely by the thermal spectrum of the CMB target
photons \cite{Stecker}.

\noindent\hangindent=25pt {\bf $I$} \hspace{.5cm} Or {\bf $I(>E)$}.  This is
the integral intensity (cosmic rays per unit area per unit solid angle
per unit time). ``Integral intensity'' means integrated over all
energies above $E$.

\noindent\hangindent=25pt {\bf $t$} \hspace{.5cm} The total operating
time of the observatory, measured in years.

\noindent\hangindent=25pt {\bf $A_0$}  \hspace{.5cm} The ground area of the
observatory measured in $km^2$.  For a source at zenith angle
$\theta$, the effective collecting area is $A=A_0 \cos \theta$. 

\noindent\hangindent=25pt {\bf $\cal E$}\hspace{.5cm} The detector's
exposure to a discrete source (and its part of the sky).  This
exposure is measured in $km^2 yr$.  For example, a detector array with
acceptance out to 60 degrees zenith angle has an exposure to a source
that passes through its zenith given by
$${\cal E} = detector\_area\times(live\_time)\times \sqrt3/2\pi.$$

\noindent\hangindent=25pt {\bf $f_D$}\hspace{.5cm} The fraction of the
sky well exposed to the detector.  For an observatory with uniform
full-sky exposure, $f_D$ would be 1.  For a detector at a single
mid-latitude site with uniform acceptance to $60^{\circ}$ zenith
angle, the acceptance varies with declination, but an effective
fraction $f_D=1/2$ is appropriate.  See section 10.

\noindent\hangindent=25pt {\bf $N$}\hspace{.5cm} The total number of
arrival directions in a cosmic ray data set from a single observatory
site.  For the case of a full-sky observatory with uniform celestial
exposure, $N$ is 1/2 the total number of events in the data set.

\noindent\hangindent=25pt {\bf $\omega$}\hspace{.5cm} The solid angle
over which the signal events arrive from a particular source.  These
may be expected to be distributed as a 2-dimensional Gaussian given by
width $\sigma$.  In that case, the solid angle is given by
$\omega=4\pi \sigma^2$.

\noindent\hangindent=25pt {\bf $\cal N$}\hspace{.5cm} Noise, i.e. the
amount of fluctuation in the expected background count.  For high
statistics, the noise is the square root of the background:
$${\cal N}=\sqrt{I \omega \cal{E}}.$$

\noindent\hangindent=25pt {\bf $\cal S$}\hspace{.5cm} Signal.  This might
be the number of counts in a target region above an expected
background count.  If a source is expected to produce a Gaussian
distribution of arrival directions, then the signal $\cal S$ might be the
Gaussian-weighted sum (minus the expected Gaussian-weighted sum from
the isotropic background).

\noindent\hangindent=25pt {\bf $\Sigma$}\hspace{.5cm} Signal-to-noise
detection threshold.  For example, if a 5-sigma detection is required,
$\Sigma$ would be set equal to 5.  It is the number of sigmas deemed
necessary to qualify for a positive detection.  In a prescribed single
trial, $\Sigma=3$ might be appropriate \cite{Clay}.

\noindent\hangindent=25pt {\bf $K$}\hspace{.5cm} The number of
in-target arrival directions needed for a background-free positive
source detection.  For example, a cluster of 5 super-GZK arrival
directions might constitute persuasive evidence for a source in a
full-sky survey, so K might be set to 5.  For a previously identified
most promising candidate source, finding even 2 or 3 new arrival
directions in a small target region might be sufficient evidence.  The
number adopted for K depends on the search circumstances.

\noindent\hangindent=25pt {\bf $\kappa$}\hspace{.5cm} The angular
diffusion coefficient that governs how the solid angle of magnetic
blurring increases with source distance: $\omega=\kappa r$.

\noindent\hangindent=25pt {\bf $B$}\hspace{.5cm} Extragalactic
magnetic field strength measured in nanogauss.

\noindent\hangindent=25pt {\bf $Z$}\hspace{.5cm} Electric charge of
the cosmic rays, measured in units of the proton charge.

\noindent\hangindent=25pt {\bf $\rho$}\hspace{.5cm} Larmor radius.  To
a good approximation, $\rho=E/ZB$, where B is the transverse field
strength in $nG$, energy $E$ is measured in $EeV$, $Z$ is electric
charge, and $\rho$ is measured in $Mpc$.  This formula is appropriate
for estimations in intergalactic space.  Within the Galaxy, the same
formula can be used if $\rho$ is measured in $kpc$ and $B$ is measured
in $\mu G$.  See Appendix B.

\noindent\hangindent=25pt {\bf $L$}\hspace{.5cm} The magnetic field
coherence length measured in $Mpc$.  This is a typical distance over
which an intergalactic magnetic field affecting ultra-high energy
cosmic rays can be regarded as having a consistent direction.

\noindent\hangindent=25pt {\bf $\mu$}\hspace{.5cm} The expected number
of cosmic ray sources within a volume of specified radius.  The radius
is usually the distance limit at which any source is expected to be
detectable.  

\noindent\hangindent=25pt {\bf $\Omega_J$}\hspace{.5cm} The emission
solid angle of a single jet (collimated emission of cosmic rays).

\noindent\hangindent=25pt {\bf $\Omega$}\hspace{.5cm} The solid angle
on a sphere at distance $r$ from a source over which the particles
from a jet have spread.  

\noindent\hangindent=25pt {\bf $\eta$}\hspace{.5cm} The spacetime
density of bursting sources (rate of bursts per unit volume).  For an
accumulation time $T$, $n\equiv \eta T$ is the spatial density of
bursts that contributed to the present-day cosmic rays, and
$R=1/n^{1/3}$ is the mean separation between the relic fossils of those
bursts.

\noindent\hangindent=25pt {\bf $W$}\hspace{.5cm}  The number of cosmic
rays emitted in a burst.  For identical sources of bursts, this number
is the same for all of them.

\noindent\hangindent=25pt {\bf $\tau$}\hspace{.5cm} The time interval
over which cosmic ray protons from an instantaneous burst arrive at
Earth.  The spread in time is caused by different trajectory lengths
due to intervening magnetic fields.  The time interval $\tau$ is
expected to increase with distance to the source and decrease with
cosmic ray energy.

\noindent\hangindent=25pt {\bf $\alpha$}\hspace{.5cm} The coefficient
that governs how the time spread $\tau$ of received cosmic rays from a burst
source increases with distance to the source: $\tau = \alpha r$
(cf. Appendix C).

\noindent\hangindent=25pt {\bf $\xi$}\hspace{.5cm} Used in Appendix A,
$\xi$ is a dimensionless measure of distance given by $\xi=r/cT$.

\section{The basic model}

The simplest model is that the observed cosmic ray intensity $I(>E)$
is the result of isotropic emission by sources with identical
luminosity $Q$ and spatial density $n$ over some history of $T$ years.
The density of cosmic rays is then given in terms of the sources by
$nQT$.  That same cosmic ray density is given in terms of the observed
intensity as $4\pi I/c$.  Equating these two expressions yields the
source luminosity
\begin{equation}\label{basic:luminosity}
Q=\frac{4\pi I R^3}{cT},\end{equation}
where the density $n$ has been
expressed in terms of the mean source separation $R$ such that
$n=1/R^3$.  The flux from a source at distance $r$ is $Q/4\pi r^2$,
and multiplying this by the exposure $\cal E$ gives the signal
\begin{equation}\label{basic:signal}
{\cal S}=\frac{IR^3}{cTr^2}\cal E.\end{equation} Exposure $\cal E$ is
the time-integrated perpendicular collecting area for flux coming from
the source direction, and $\cal E$ is measured in units of $km^2 yr$.

Note: The flux $Q/4\pi r^2$ should be multiplied by $e^{-r/cT}$ if the
effective accumulation time $T$ in equation \ref{basic:luminosity} is governed by a propagation attenuation such as GZK pion photoproduction.  
Including this factor prevents a simple analytic solution.  The
factor is irrelevant for the sub-GZK analysis.  Omitting the factor
leads to a valid result also in the super-GZK case in typical
circumstances, as explained in Appendix A.

The product $I\cal E$ is closely related to the total number of events
in the data set.  That number is a convenient parameter.
Consider first a single observatory site such as Auger
South, and let $N$ be the number of events above some energy cut.
To be specific, suppose this data set includes those events with zenith
angles less than $60^\circ$.  For a source that passes through the
zenith of the detector, it is observable 1/3 of the operating time and
the average collecting area is 
$$<A>=\frac{3}{\pi}\int_{0}^{\pi/3}A_0 cos(\theta)
d\theta=\frac{3\sqrt{3}}{2\pi}A_0$$ where $A_0$ is the full detector
area and $\theta$ is zenith angle.  The exposure to the source is then
${\cal E}=\frac{t}{3}<A>=\frac{\sqrt{3}}{2\pi}A_0\ t$, where $t$ is
the cumulative detector operating time.  The number of events $N$ in
the full data set is the cosmic ray intensity $I$ times the product of
aperture and operating time $t$.  The detector's aperture (accepting
events out to $60^{\circ}$ zenith angle) is $\frac{3\pi}{4}A_0$, so
$N=\frac{3\pi}{4}A_0\ I\ t$.  Combining this with the expression for
$\cal E$ yields
$$I{\cal E}=\frac{2\sqrt{3}}{3\pi^2}\ N=0.117\ N.$$ For a full-sky
observatory (a second site in the other hemisphere), the exposure to
this source would be little changed, but the number of events in the
data set would be twice as great.  Using the approximation that a
two-site observatory has uniform celestial exposure \cite{Sommers},
the calculation for the one source here would apply to all sources.
The rule to be adopted, therefore, is that $I{\cal E}=0.117\ N$, where
$N$ is the number of events in the data set for a single-site
observatory or half the number of events in the data set of a full-sky
observatory.  (See section 10 for further discussion of non-uniform
celestial exposure.)\\

\noindent {\it \underline{Super-GZK analysis:}}\\

Suppose $K$ signal showers are needed to make a positive source
detection in a signal-limited regime.  Substituting $K$ for $S$ in
equation \ref{basic:signal} gives

\begin{equation}\label{basic:sup:rK}
K=\frac{I{\cal E}R^3}{cTr^2} \Rightarrow r_K^2=\frac{I{\cal E}R^3}{cTK}
\end{equation}
where $r_K$ is the distance at which the expected signal $\cal S$
would be equal to $K$.  A source is detectable within that radius
around us if it is in the fraction $f_D$ of the celestial sphere.  The
expected number $\mu$ of detectable sources is that exposed volume times the 
source density $1/R^3$:
$$\mu = \frac{4\pi f_D}{3} r_K^3 /R^3.$$
Substituting the previous expression for $r_K$, this becomes
\begin{equation}\label{basic:sup:mu} 
\mu=\frac{4\pi f_D}{3}\ (\frac{I{\cal E}R}{cTK})^{3/2}.
\end{equation}
Denote by $r_0$ the distance to
the nearest source.  What condition on $R$ ensures that $r_0$ is
likely to be less than $r_K$?  The probability that $r_0$ is less than
$r_K$ is the complement of the Poisson probability that there are 0
sources within that volume when the expected number is $\mu$:
$$P(r_0< r_K)=1-Poiss(0;\mu)=1-e^{-\mu}.$$
Setting this final expression to be greater than 1/2 (i.e. the
condition that detection is more likely than not), gives the condition
$\mu>ln(2)$.  Then substituting the above formula for $\mu$ yields 
this lower limit for the mean separation of sources:
\begin{equation}\label{basic:sup:R1}
R>(\frac{3 ln 2}{4\pi f_D})^{2/3}\ \frac{cTK}{I\cal E}.
\end{equation}
Alternatively, using $I{\cal E}=0.117 N$ and $f_D=1/2$, the minimum separation
  $R_{min}$ for which the nearest source is likely to be detected is
  given by 
\begin{equation}\label{basic:sup:R2}
\boxed{R_{min}=4.1\ \frac{K}{N}\ cT.}
\end{equation}

This equation quantifies the conditions for detectability in the
signal-limited regime.  The required minimum separation $R_{min}$
shrinks for any energy cut (fixed $T$) as the number of arrival
directions $N$ increases (e.g. as an observatory's exposure increases
over time).

The expression for $r_K$ in equation \ref{basic:sup:rK} is derived
without including GZK flux attenuation.  The resulting expressions for
$\mu$ and $R$ are valid only if $r_K$ is much smaller than $cT$.  As
discussed in Appendix A, this is usually satisfied.  The above
expression for $R_{min}$ is therefore normally justified.  There is
also an {\it upper} limit on $R$, since the nearest source should not be
much farther away than the attenuation length $cT$.  Appendix A shows
how to find this upper limit $R_{max}$ for which detection of the
nearest source is likely.\\

\noindent {\it \underline{Sub-GZK analysis:}}\\

The signal from a single source is presumed to be spread over a solid
angle $\omega$ by virtue of small magnetic bends while passing through
intergalactic space with some unknown spectrum of magnetic field
strengths and randomly changing directions.  By invoking the central
limit theorem, or by analogy with multiple Coulomb scattering, it is
natural to postulate that this magnetic blurring increases linearly
with distance: $\omega=\kappa r.$ This relation defines the diffusion
coefficient $\kappa$.  (The RMS angle in any plane that contains the
central direction to the source increases with the square root of the
distance traveled.  The solid angle increases linearly with distance
because it is proportional to the product of two such angles.)

The expected background within the solid angle $\omega$ is $I\omega \cal
E.$  The noise is the fluctuation in this background.  For large
statistics, that is the square root of the background, so 
$${\cal N}=\sqrt{I\kappa r \cal E}.$$
Using the expression for $\cal S$ above, the signal-to-noise ratio is
$${\cal S}/{\cal N}=\sqrt{\frac{I {\cal E}}{\kappa}}\frac{R^3}{cTr^{5/2}}.$$
Now suppose a detection requires that ${\cal S}/{\cal N}>\Sigma$ for some number
$\Sigma$, and denote by $r_{\Sigma}$ the distance at which this
signal-to-noise ratio occurs.  The equation above gives
\begin{equation}\label{basic:sub:rsigma}
r_{\Sigma}^{5/2} = \sqrt{\frac{I \cal E}{\kappa}}\frac{R^3}{cT\Sigma}.
\end{equation}

A source is detectable within the radius $r_{\Sigma}$ around us.  
The expected number of sources within that radius is the volume times
the source density:
\begin{equation}\label{basic:sub:mu}
\mu = \frac{4\pi f_D}{3} r_{\Sigma}^3/R^3\ \ \Leftrightarrow\ \
\mu=\frac{4\pi f_D}{3}\ (\frac{I{\cal E}R}{\kappa})^{3/5}\
(\frac{1}{cT\Sigma})^{6/5}.
\end{equation}
 Denoting by $r_0$ the distance to the
nearest source, as above, detection of the nearest source is likely if
the probability is greater than 1/2 that $r_0$ is less than
$r_{\Sigma}$.  That probability is given by the complement of the
Poisson probability that there are 0 sources within that volume when
the expected number is $\mu$.  As above, this means $\mu>ln(2)$, which
here reduces to
\begin{equation}\label{basic:sub:R1}
R\ >\ (\frac{3 ln 2}{4\pi f_D})^{5/3}\ (cT\Sigma)^2\
\frac{\kappa}{I \cal E}.
\end{equation}
This is the constraint on the mean source separation $R$ such that a
detection of significance $\Sigma$ is likely.

Using $I{\cal E}=0.117 N$ and $f_D=1/2$, the minimum source separation for which it
is likely that the nearest source is detectable becomes
\begin{equation}\label{basic:sub:R2}
\boxed{R_{min}=1.4\ (cT\ \Sigma)^2\ \frac{\kappa}{N}.}
\end{equation}

The smearing by magnetic deflection is presumed to be the result of
many small-angle scatterings as a charged particle passes through
irregular magnetic fields between the source and the Earth.  The
process is mathematically similar to multiple Coulomb scattering, and
a Gaussian distribution of arrival directions can be expected.  In
that case, the solid angle $\omega$ is not a simple target region.
Instead, each arrival direction should be weighted in proportion to
the expected Gaussian distribution.  The simple $S/N$ analysis here
still pertains, provided the weighting factor is taken to be $4\pi
\sigma^2$ times the Gaussian probability distribution.  See Appendix D
for further details.

\section{Fiducial calculations}

The inequalities \ref{basic:sup:R1}, \ref{basic:sub:R1} and equations
\ref{basic:sup:R2}, \ref{basic:sub:R2} give expressions for the
minimum value of the mean source separation $R$ such that the nearest
source is likely to be detectable.  The minimum mean separations
depend on variables that are not known with certainty.  Here some
fiducial parameter values are adopted for illustration.  Readers who
favor other parameter values can readily scale the answers for their
values.  

For the super-GZK estimate, the fiducial estimates here will be based
on $cT=100\ Mpc$ and $N=50$.  There is some energy for which the
accumulation time is $(100\ Mpc)/c$, and any observatory should
eventually detect 50 air showers above that energy.  For $K$, it will
be assumed here that there is reason to suspect the existence of a
source at a particular location and that a single-trial test has been
prescribed.  A cluster of $K=3$ arrival directions at that celestial
position would be a strong positive result if there are only 50 events
in a data set covering much of the sky.

With these adopted values for the parameters $K$, $cT$, and $N$,
equation \ref{basic:sup:R2} gives $R_{min}\approx 25\ Mpc$.  The
nearest source is likely to be detectable if the sources are separated
by more than 25 $Mpc$ on average.

A fiducial calculation for the sub-GZK case requires adopting a value
for $\kappa$, the coefficient that controls how magnetic blurring
increases with source distance.  Wild guesses are allowed here as the
properties of extragalactic magnetic fields are poorly determined by
observations or theory \cite{Magfields}.  One simplistic model is that
the magnetic field has a typical strength $B$ that is randomly
oriented, but a particle experiences the same orientation for
coherence length L.  Its total path of length $r$ is made up of $r/L$
deflections from these randomly oriented fields.  Appendix B shows
that, in this simplistic model, $\kappa$ is given by

\begin{equation}\label{kappa}
\kappa = \frac{4\pi}{9}\ LZ^2B^2/E^2.
\end{equation}
 If $L=1\ Mpc,\ B=1\ nG,\
Z=1,$ and $E=10\ EeV$, then $\kappa=\frac{\pi}{900}=0.014\ sr/Mpc.$ 

As an example, suppose $cT=1500\ Mpc$ (a typical survival distance for
nucleons against pair production at energies above the spectrum's
ankle), and let $N=10^4$ arrival directions.  As in the
super-GZK fiducial estimate, suppose a prescribed test has been
applied to a suspected celestial location, so $\Sigma=3$ is
statistically significant (a 3-sigma result for a single trial).
Using $cT=1500\ Mpc$, $N=10,000$, $\Sigma=3$, and $\kappa=0.014$,
equation \ref{basic:sub:R2} gives $R_{min}=40\ Mpc$.  The nearest
source is expected to be detectable at this sub-GZK energy provided
the sources are separated by at least 40 $Mpc$ on average.

The estimate for the detection distance $r_{\Sigma}$ would not be
meaningful if the magnetic blurring at that distance (and for the
energy cut $E$) were to produce a solid angle greater than $2\pi$.  In
this fiducial calculation with $\kappa=0.014\ sr/Mpc$, the magnetic
blurring for a source at 40 $Mpc$ is 0.56 $sr$, corresponding to a
circle of 24 degrees radius on the sky.

The fiducial parameters adopted here may be quite wrong.  They are
presented only to illustrate the use of the formulas and provide
explicit answers that can be easily scaled for different values of the
parameters.

\section{Comments about the dependences}

In the super-GZK regime, the signal can be maximized by lowering the
energy cut as much as possible without violating the signal-limited
condition.  For the actual mean source separation R, therefore,
detecting the nearest source is made more probable by reducing the
energy cut as much as possible in the signal-limited regime.

In the sub-GZK regime, there may be little dependence of
$R_{min}$ on the energy cut.  At least for the simplistic model
treated in Appendix B, $\kappa$ depends on energy as $1/E^2$, which is
approximately the same as the energy dependence of the intensity
$I(>E)$ or number of events $N$.  The factor $\frac{\kappa}{I\cal E}$
is almost independent of energy, provided the detector's acceptance
(aperture) is not growing with energy.  Increasing the minimum energy
reduces the magnetic blurring, but the effect on $S/N$ is offset by
the lower statistics.  (There may be some advantage in raising the
minimum energy in the analysis if the detector's acceptance does
increase with energy.)  Near the GZK threshold, energy dependence
enters also through the accumulation time T.  Raising the minimum
energy in the analysis would then reduce $R_{min}$.

The statistical criterion for detection encoded in $\Sigma$ enters
quadratically in the inequality \ref{basic:sub:R1} and equation
\ref{basic:sub:R2}.  As noted above, this number can be made
relatively small if a careful sky survey determines a single best
candidate source, its solid angle extent, and an optimal energy cut.
A 3-sigma result for a single test with new data would then be a
compelling result.

Intergalactic magnetic fields might someday be studied using discrete
sources of cosmic rays at known distances.  For now, gross uncertainty
is encoded in the single coefficient $\kappa$, and any fiducial
calculation based on an adhoc value of $\kappa$ should be treated with
appropriate suspicion.

Large detector exposure $\cal E$ is obviously crucial for detecting
discrete sources.  Inequalities \ref{basic:sup:R1} and
\ref{basic:sub:R1} show that the minimum value for the mean
distance between sources in order for them to become detectable will
shrink inversely as the exposure increases.  A vast increase in
exposure would lead to source detections, or else it would radically
shrink the viable parameter space that describes the source
distribution and magnetic fields.  See section 10 for additional
discussion about the dependence on exposure.

The expected number of detectable sources $\mu$ grows in proportion to
${\cal E}^{3/2}$ for the super-GZK case of signal-limited detection.
A 10-fold increase in exposure results in a 30-fold increase in the
number of detectable sources, and this is true even if that number is
less than 1!  Success in charged particle astronomy is all about
achieving huge exposure.

The ${\cal E}^{3/2}$ dependence of $\mu$ on $\cal E$ is closely
related to the ``logN-logS'' relation of astronomy, in which the
source count varies like $flux^{-3/2}$ for a homogeneous distribution
of detectable sources in Euclidean space, independent of their
luminosity distribution.  Since $flux=K/{\cal E}$ in the analysis
here, the number of detectable sources increases with exposure like
${\cal E}^{3/2}$.

For background-limited (sub-GZK) analysis, the number of sources grows
with exposure also, but it increases less rapidly with ${\cal E}$.
The number of detectable sources is proportional to ${\cal E}^{3/5}$.

\section{Sources with jets}

Jets are a common phenomenon among objects known to produce
energetic particles.  Relativistic bulk plasma motion is advantageous
in accelerating particles to high energies, so it is reasonable to
conjecture that the highest energy cosmic rays could be emitted in
collimated jets.  The previous assumption of isotropic emission from
cosmic ray sources might be inappropriate.

Suppose a single jet emits ultra-high energy cosmic rays in a solid
angle $\Omega_J$.  The
luminosity of every source is calculated as before: 
$$Q=\frac{4\pi I R^3}{cT}.$$ 

The measured signal from any one source is zero unless the
observer is within the solid angle of the beam, which grows with
distance $r$ from the source by magnetic deflections.  Let $\Omega$
be the solid angle on the sphere of radius $r$ which gets flux from
the jet.  Then the signal S is
$${\cal S}=\frac{Q}{\Omega r^2}{\cal E} = \frac{4\pi}{\Omega} \frac{IR^3}{cTr^2}{\cal E}.$$ \\

\noindent {\it \underline{Super-GZK analysis:}}\\

A simple assumption is that the magnetic rigidity of super-GZK
particles is high enough that $\Omega \approx \Omega_J$, i.e. magnetic
deflection does not significantly increase the solid angle of the jets
in transit between the source and Earth.  The signal $S$ at distance $r$
is then given by the previous equation with $\Omega$ replaced by
$\Omega_J$.  

For a given exposure $\cal E$ and cosmic ray intensity $I$ accumulated
over $T$ years, $K$ particles would be expected at distance $r_K$:
\begin{equation}\label{jets:sup:rk}
r_K^2 = \frac{4\pi}{\Omega_J}\frac{I{\cal E}R^3}{cTK}.
\end{equation}
The probability is $\Omega_J/4\pi$ that Earth is in the beam of any
one source, so the density of viewable sources is
$\frac{\Omega_J}{4\pi R^3}$.  The expected number of {\it
detectable} sources (i.e. pointing at us) within the exposed volume of radius
$r_K$ around us is this:
\begin{equation}\label{jets:sup:mu}
\mu = \frac{f_D \Omega_J}{3} \frac{r_K^3}{R^3} =
  \frac{f_D}{3\sqrt{\Omega_J}}(\frac{4\pi I{\cal E}R}{cTK})^{3/2}.
\end{equation}

As in the basic model above, the probability that the nearest source is
closer than $r_K$ is greater than 1/2 provided $\mu>ln2$.  Using the
above expression for $\mu$ converts this inequality to a lower limit
on the mean separation of sources:

\begin{equation}\label{jets:sup:R1}
R>(\frac{\Omega_J}{4\pi})^{1/3}\ (\frac{3 ln 2}{4\pi f_D})^{2/3} \
\frac{cTK}{I\cal E}.
\end{equation}
Using $I{\cal E}=0.117N$ and $f_D=1/2$ as before, this becomes

\begin{equation}\label{jets:sup:R2}
\boxed{R_{min}=4.1\ (\frac{\Omega_J}{4\pi})^{1/3}\ \frac{K}{N}\ cT.}
\end{equation}
Detection of the nearest source is more likely than not provided the
sources have at least this mean separation.  Notice that the minimum
separation is decreased relative to the basic model by the cube root
of the jet opening solid angle (as a fraction of the isotropic
$4\pi$ solid angle).  

A fiducial value of $\Omega_J$ will here be taken to be 0.01 $sr$,
corresponding to a jet opening cone of $3.2^{\circ}$ half-angle.
Together with the previously adopted fiducial values ($cT=100\ Mpc,\
N=50,\ K=3$), the fiducial calculation here gives $R_{min}=2.3\
Mpc$, suggesting that the nearest source should be detectable even if
the sources are distributed more densely than normal galaxies.  

The minimum separation $R_{min}$ is substantially smaller than in the
basic model because the signal is strong when looking into a jet.  The
total luminosity of every source can be relatively weak (as must be
the case if the density of sources is high), and an individual source
can nevertheless be detected far away.  (For $R_{min}=2.3\ Mpc$ in
this fiducial estimate, the maximum detection distance $r_K$ is $17\
Mpc$, but that detection distance grows in proportion to $R^{3/2}$).

In the super-GZK regime, jet sources are easier to detect than
isotropic sources if the mean separation between sources is small, but
they are more difficult to detect if the mean source separation is
large.  Since $r_K/R$ is larger for jet sources than for isotropic
sources, it may happen that there are many sources within the GZK
volume but none of them is pointing at us.  For example, suppose
$R=25\ Mpc$ (the minimum mean separation for detectability in the
fiducial estimate for the isotropic case).  The expected number of
sources within 100 $Mpc$ of Earth in the exposed half of the sky would
be 134.  For $\Omega_J=0.01\ sr$, the probability is only
$\frac{0.01}{4\pi}=8\times 10^{-4}$ for Earth to be in the beam of any
one source.  Although there are 134 sources readily detectable if any
points at us, the expected number pointing at us is only 0.1.
Detection is unlikely for a large mean source separation.  \\

\noindent {\it \underline{Sub-GZK analysis:}}\\

Although super-GZK particles can be assumed to maintain their
directions in transit from source to Earth, that is not expected in
the sub-GZK regime.  As sub-GZK particles get farther from the source,
the effective angular extent of the jet is dominated by their magnetic
deflections rather than the emission angle $\Omega_J$ of the jet
itself.  Here it is therefore assumed that, at the distance to Earth,
the flux from a single jet is spread over a solid angle
$\Omega=3\kappa r$.  (See Appendix B for an explanation of the
factor 3.)  The signal at distance $r$ becomes
(cf. equation \ref{basic:signal})
$${\cal S}=\frac{4\pi}{3}\frac{I\cal E}{cT\kappa}\frac{R^3}{r^3}.$$
The background does not care that the luminosity of individual sources
is in jets.  It is given, as in the isotropic case, by $I\omega \cal
E$.  Its square root gives the estimated fluctuation in the
background.  Substituting $\kappa r$ for $\omega$, that noise is
$${\cal N}=\sqrt{I\kappa r \cal E}.$$ The signal-to-noise is then
$${\cal S}/{\cal N} = \frac{4\pi}{3}\sqrt{\frac{I\cal
    E}{\kappa^3}}\frac{R^3}{cTr^{7/2}}.$$

The analysis proceeds as in the case of isotropic sources.  Adopting a
value $\Sigma$ for ${\cal S}/{\cal N}$ gives an expression for $r_{\Sigma}$:
\begin{equation}\label{jets:sub:rsigma}
r_{\Sigma}^{7/2} = \frac{4\pi}{3}\sqrt{\frac{I\cal
    E}{\kappa^3}}\frac{R^3}{cT\Sigma}.
\end{equation}
 This is the radius at which a
    source should be detectable with signal-to-noise equal to
    $\Sigma$.

The expected number of detectable sources (beamed at us from the
exposed fraction of sky $f_D$) within the
radius $r_{\Sigma}$ is 
$$\mu = f_D \int_0^{r_\Sigma} \Omega r^2 n dr = f_D \int_0^{r_\Sigma} 3\kappa
r^3 \frac{1}{R^3} dr = f_D \frac{3}{4}\kappa r_{\Sigma}^4/R^3.$$
Using the foregoing expression for $r_{\Sigma}$, the expected number
of sources closer than $r_{\Sigma}$ becomes
\begin{equation}\label{jets:sub:mu}
\mu= \frac{3f_D}{4}(\frac{4\pi}{3})^{8/7}\ (I{\cal E})^{4/7}\
(\frac{1}{\kappa})^{5/7}\ (\frac{1}{cT\Sigma})^{8/7}\ R^{3/7}.
\end{equation}
As in the case of isotropic sources, the probability that the nearest
source is detectable (closer than $r_{\Sigma}$) is given by
$\mu>ln(2)$, which here reduces to
\begin{equation}\label{jets:sub:R1}
R\ >\ (\frac{4\
  ln2}{3f_D})^{7/3}(\frac{3}{4\pi})^{8/3}(\frac{cT\Sigma}{\sqrt{I\cal
  E}})^{8/3}\kappa^{5/3}.
\end{equation}

Substituting $0.117N$ for $I{\cal E}$ and 1/2 for $f_D$ as previously,
this becomes 
\begin{equation}\label{jets:sub:R2}
\boxed{R_{min}=1.6(\frac{\Sigma^2}{N})^{4/3}(cT)^{8/3}\kappa^{5/3}.}
\end{equation}

Using again the fiducial parameter values $\Sigma=3,\ cT=1500\ Mpc,\
N=10,000,\ \kappa=0.014\ sr/Mpc$ for sub-GZK analysis, this formula
gives $R_{min}=33\ Mpc$.  With this minimum source separation, sources
are detectable ($\Sigma>3$) out to $r_{\Sigma}=47\ Mpc$, and the
arrival directions are spread over solid angle $\omega=0.66\ sr$
($26^{\circ}$ cone half-angle).

The sub-GZK regime offers detection capability for jetted sources with
large mean separation even though super-GZK detection might be
unlikely in that case.  The two regimes are complementary.  Super-GZK
astronomy is likely for densely distributed sources with jets, whereas
sub-GZK astronomy is likely for more widely distributed sources with
jets.

\section{Isotropic burst sources}

It is unlikely that the sources of high energy cosmic rays are
permanent.  The universe is dynamic, and even the most powerful active
galactic nuclei may be temporary feeding episodes of supermassive
black holes.  For temporary sources, the source density $n$ should be
interpreted as the mean density of {\it active} sources at any 
time.  The formulas of the preceding sections should then be
appropriate.

Some modifications are needed, however, if cosmic rays are produced in
brief bursts.  Due to the magnetic wandering of particles en route to
the Earth, the duration of flux from an instantaneous burst increases
with distance from the source.  The duration $\tau$ (at any cosmic ray
energy) should increase linearly with distance: $\tau=\alpha r$ (where
$\alpha$ is energy dependent, {\it cf.} Appendix C).  In this paper,
cosmic ray sources are regarded as ``bursts'' if their emission
lifetimes are not long compared to the time spread $\tau$ expected at
Earth for a source that emits all of its cosmic rays instantaneously.
Let $W$ be the magnitude of a burst, i.e. the total number of emitted
cosmic rays.  If an isotropic burst is at distance $r$ from the Earth,
the flux over time $\tau$ is $W/(4\pi r^2 \tau)$.  Substituting
$\alpha r$ for $\tau$ and multiplying by the exposure $\cal E$ gives
the signal,
$${\cal S}=\frac{W{\cal E}}{4\pi \alpha r^3}.$$

Let $\eta$ denote the spatial density of bursts per unit time.  If $T$
is the accumulation time for cosmic rays, then $n\equiv\eta T$ is the
fossil density of bursts that have contributed to the accumulated cosmic
ray density, so
$$\frac{4\pi}{c}I = nW = \eta T W.$$
Solving this for $W$ and inserting the result into the expression for
the signal $\cal S$ gives
$${\cal S} = \frac{I\cal E}{\alpha c\ \eta T\ r^3 }.$$

An estimate for $\alpha$ derived in Appendix C is
$$\alpha=\frac{L^2Z^2B^2}{36 c E^2}.$$

\noindent {\it \underline{Super-GZK analysis:}}\\

As usual, suppose K events are needed for a detection:
\begin{equation}\label{isoburst:sup:rK}
{\cal S}=K\ \Rightarrow\ r_K^3=\frac{I\cal E}{\alpha c\ \eta T\
  K}.
\end{equation}
The density of burst sources at distance $r$ with flux now ``on'' is
$\eta \tau=\eta \alpha r$.  The number of detectable bursts closer
than $r_K$ is therefore
$$\mu = f_D \int_0^{r_K}4\pi r^2\ \eta\ \alpha r\ dr = \pi f_D \alpha \eta
r_K^4.$$  Using the foregoing expression for $r_K$ and $R^3=1/(\eta
T)$, this can be written as
\begin{equation}\label{isoburst:sup:mu}
\mu=\frac{\pi f_D}{cT}\ (\frac{1}{\alpha c})^{1/3}\ (\frac{I\cal
  E}{K})^{4/3}\ R.
\end{equation}

The probability $P(r_0<r_K)$, that the nearest (on) source is closer
than $r_K$, is greater than 1/2 (detection likely) provided
\begin{equation}\label{isoburst:sup:R1}
\mu>ln(2)\ \ \Leftrightarrow\ \ R > \frac{ln 2}{\pi f_D}\ cT\ (\alpha c)^{1/3}\ (\frac{K}{I\cal E})^{4/3}.
\end{equation}

Substituting $0.117N$ for ${I\cal E}$ and 1/2 for $F_D$ as before, this
becomes 
\begin{equation}\label{isoburst:sup:R2}
\boxed{R_{min}=\frac{3.9}{f_D}(\alpha c)^{1/3}(\frac{K}{N})^{4/3} cT.}
\end{equation}

A fiducial calculation is obtained from the parameters used previously
for super-GZK estimates together with $\alpha c=\frac{1}{36E^2}$ (and
$E=100\ EeV$).  These values give $R_{min}=0.26\ Mpc$.  This minimum
$R$ corresponds to a maximum fossil density $\eta T=1/R^3=60/Mpc^3$
accumulated over $T\approx 3\times 10^8\ yrs$ ($cT=100\ Mpc$).  For
the average galaxy density $0.01/Mpc^3$, this fossil density requires
a burst frequency not more than one per 50,000 years per galaxy.  (The
minimum mean time between bursts per galaxy is $<\Delta t>_{min}=0.01
R^3T$.)\\

Note, however, the strong dependence of this minimum burst interval on
the size $N$ of the data set, it being proportional to $R^3\sim 1/N^4$.
A five-fold increase in exposure would reduce the limit of 50,000
years down to 80 years.  Detection of a discrete source would then be
likely provided the mean time interval between bursts is not greater
than 80 years in a typical galaxy.  Increasing exposure is especially
advantageous in the search for discrete sources if the sources are
isotropic bursts.\\

\noindent {\it \underline{Sub-GZK analysis:}}\\

The background noise is not changed by the assumption of burst sources.  It
is still given by $${\cal N}=\sqrt{I\omega \cal E} = \sqrt{I\kappa r \cal
  E}.$$
The signal-to-noise is therefore
$${\cal S}/{\cal N} = \frac{I\cal E}{\alpha c\ \eta T\ r^3 }\ /\
\sqrt{I\kappa r \cal E} = \sqrt{\frac{I\cal E}{\kappa}}\
\frac{1}{\alpha c\ \eta T\ r^{7/2}}.$$ For a specified value $\Sigma$
for $S/N$, the maximum detection distance is $r_{\Sigma}$.  Using
$\eta T=1/R^3$, $r_{\Sigma}$ is given by
\begin{equation}\label{isoburst:sub:rsigma}
r_{\Sigma}^{7/2} = \sqrt{\frac{I\cal E}{\kappa}}\frac{R^3}{\alpha c
    \Sigma}.
\end{equation}
The expected number of sources closer than this distance is 
$$\mu = f_D\int_0^{r_{\Sigma}}4\pi r^2\ \eta \tau\ dr = f_D
\int_0^{r_{\Sigma}}4\pi \alpha \eta r^3 dr = \pi f_D \alpha \eta
r_{\Sigma}^4,$$
which, upon substituting for $r_{\Sigma}$ becomes
\begin{equation}\label{isoburst:sub:mu}
\mu=f_D \frac{\pi}{cT}\ (\frac{1}{\alpha c})^{1/7}\ \ (\frac{I\cal
  E}{\kappa \Sigma^2})^{4/7}\ \ R^{3/7}.
\end{equation}
Detection is likely, i.e. $Prob(r_0<r_{\Sigma})>1/2$, if $\mu>ln(2)$,
which is
\begin{equation}\label{isoburst:sub:R1}
R > (\frac{ln 2}{\pi f_D})^{7/3}\ (cT)^{7/3}\ (\alpha c)^{1/3}\
(\frac{\kappa \Sigma^2}{I\cal E})^{4/3}.
\end{equation}

Using the estimate $I{\cal E}=0.117N$, this becomes
v\begin{equation}\label{isoburst:sub:R2}
\boxed{R_{min}=0.51(\frac{cT}{f_D})^{7/3} (\alpha c)^{1/3}
(\frac{\kappa \Sigma^2}{N})^{4/3}.}
\end{equation}

A fiducial estimation can be done using $\alpha c=1/(36E^2)$ (for
$E=10\ EeV$) and the same parameters that were used for steady
isotropic sources: $cT=1500\ Mpc,\ \kappa = 0.014\ sr/Mpc,\ \Sigma=3,\
N=10,000, f_D=1/2.$ This gives $R_{min}=1.3\ Mpc$.  This minimum $R$
corresponds to a maximum fossil density $\eta T=1/R^3=0.5/Mpc^3$
accumulated over $T\approx 4.5\times 10^9\ yrs$ ($cT=1500\ Mpc$).  For
the average galaxy density $0.01/Mpc^3$, this fossil density requires
a burst frequency not more than one per $9.0\times 10^7$ years per
galaxy.

As in the super-GZK case, this minimum mean time between bursts in
galaxies is proportional to $1/N^{4}$.  Modest increase in exposure
can dramatically reduce the minimum time between bursts that would make
the detection of a burst source likely.

\section{Jet bursts}

The previous two sections considered variations in which the cosmic
ray sources emit in collimated jets or emit in isotropic bursts.  It
could also be that cosmic rays are emitted in collimated bursts.  Both
types of modifications to the basic model should then be incorporated
together.  The analysis here follows the now-familiar progression.

Each burst emits a total number $W$ of cosmic rays above some energy
threshold, and they are emitted into a solid angle $\Omega_J$.  The
spacetime density of bursts is $\eta$ bursts per unit volume per unit
time.  For an accumulation time $T$, the spatial density of bursts
that contributed to the present cosmic ray intensity is $n\equiv\eta
T$.  Thus, 
$$\frac{4\pi}{c}I = nW = \eta T W\ \ \Rightarrow\ \ W =
\frac{4\pi}{c}\frac{I}{\eta T}.$$
The duration of a burst at distance $r$ from the source is $t=\alpha
r$, and a detector with exposure $\cal E$ to that part of the sky
will collect signal $\cal S$ if it is within the solid angle $\Omega$ of the
(magnetically spreading) jet, where
$${\cal S}=\frac{W\cal E}{\Omega r^2 t} = \frac{W\cal E}{\Omega r^2 \alpha r}
= \frac{4\pi I\cal E}{\alpha c\ \eta T\ \Omega r^3}.$$\\

\noindent \underline{\it Super-GZK analysis:}\\

As in the earlier jet source analysis, suppose $\Omega=\Omega_J$,
i.e., the super-GZK particles retain their directions enough that the
emission angle of the jet is approximately the solid angle of the
jet's flux at the distance of Earth.  Suppose $K$ particles from the source are
required for a signal-limited (super-GZK) detection.  The distance
$r_K$ is the distance at which a source is expected to produce that
many signal events in a detector with exposure $\cal E$:
\begin{equation}\label{jetburst:sup:rK}
S=K\ \ \Rightarrow\ \ r_K^3=\frac{4\pi I \cal E}{\alpha c\ \eta T\
  \Omega_J K}.
\end{equation}

The expected number of sources within the volume of radius $r_K$ and
exposed to the detector is 
$$\mu = f_D \int_0^{r_K} \Omega_J\ r^2\ \eta\ \tau(r)\ dr = f_D \int_0^{r_K} \Omega_J\
\alpha\ \eta\ r^3\ dr = \frac{f_D}{4}\Omega_J \alpha \eta r_K^4 =
\frac{f_D}{4}\Omega_J \alpha \eta\ (\frac{4\pi I \cal E}{\alpha c\ \eta T\
  \Omega_J K})^{4/3}.$$
Using the mean separation of fossils that have contributed in time
$T$ (so $R^3=1/(\eta T)$), the expected number $\mu$ is
\begin{equation}\label{jetburst:sup:mu}
\mu=\frac{f_D}{4}(4\pi)^{4/3}\ (\frac{1}{\Omega_J\ \alpha c})^{1/3}\
(\frac{I\cal E}{K})^{4/3}\ \frac{R}{cT}.
\end{equation}
As previously, the probability of the nearest source being detectable
is greater than 1/2 provided 
$\mu>ln(2)$.  For jet bursts,
\begin{equation}\label{jetburst:sup:R1}
\mu>ln(2)\ \ \Leftrightarrow\ \ R\ >\ (\frac{1}{4\pi})^{4/3}\ (\frac{4 ln
  2}{f_D})\ (\alpha c\ \Omega_J)^{1/3}\ (\frac{K}{I\cal E})^{4/3}\ cT.
\end{equation}

Making the approximation $I{\cal E}=0.117N$ reduces this to
\begin{equation}\label{jetburst:sup:R2}
\boxed{R_{min}=\frac{1.7}{f_D} (\alpha c \Omega_J)^{1/3}(\frac{K}{N})^{4/3}cT.}
\end{equation}
A fiducial calculation can be done using the same adhoc values that
were adopted previously: $cT=100\ Mpc,\ \alpha c=1/(36\times 100^2),\
\Omega_J=0.01\ sr,\ K=3,\ N=50,\ f_D=1/2$, giving
$R_{min}=0.024\ Mpc$.  This is the mean separation of fossils of
bursts that contributed to the cosmic ray density during the
accumulation time $T$.  That is, $R=\eta T$.  Using again the density
of 0.01 $galaxy/Mpc$, the maximum mean time between bursts in each
galaxy is given by $<\Delta t>_{min}=0.01 R^3 T$, and for $R=0.024\
Mpc$, this gives $<\Delta t>_{min}=41\ yrs$.  For the fiducial
parameter values adopted here, the nearest source is likely to be
detectable provided the mean interval between bursts in each galaxy is
at least 41 years.   \\

\noindent \underline{\it Sub-GZK analysis:}\\

As in the persistent jet sources, the solid angle of the jet is not
approximately constant except at the highest energies.  For the
background-limited analysis, it increases with distance (cf. Appendix
C): $\Omega=3\kappa r$.  Substituting this into the expression for the
jet burst signal gives
$${\cal S}=\frac{4\pi}{3}\ \frac{I \cal E}{\alpha c\ \eta T\ \kappa r^4}.$$

As usual, the background is $I\omega \cal E$, and the noise is its
square root: ${\cal N}=\sqrt{I\kappa r \cal E}$ (using $\omega=\kappa r$).
The signal-to-noise ratio is therefore
$${\cal S}/{\cal N}=\frac{4\pi}{3} \sqrt{\frac{I\cal E}{\kappa^3}}\ \frac{1}{\alpha c\ \eta
  T\ r^{9/2}}.$$
Setting a detection threshold ${\cal S}/{\cal N}=\Sigma$ gives the maximum distance
  $r_{\Sigma}$ at which this signal-to-noise ratio is expected:
\begin{equation}\label{jetburst:sub:rsigma}
{\cal S}/{\cal N}=\Sigma\ \ \Leftrightarrow\ \
r_\Sigma^{9/2}=\frac{4\pi}{3}\sqrt{\frac{I\cal E}{\kappa^3}}
\ \frac{1}{\alpha c\ \eta T\ \Sigma}.
\end{equation}

The mean density of jet bursts at distance $r$ which are ``on'' and
pointed at us is
$f_D \eta \tau \Omega/4\pi$.  Using $\tau=\alpha r$ and $\Omega=3\kappa r$, the
expected number within a volume of radius $r_{\Sigma}$ and exposed to
the detector is
$$\mu = f_D \int_0^{r_{\Sigma}} 3\ \kappa\ \eta\ \alpha\ r^4\ dr =
\frac{3f_D}{5}\ \kappa\ \alpha\ \eta\ r_{\Sigma}^5.$$
Using the foregoing expression for $r_{\Sigma}$ and $\eta T=1/R^3$,
this becomes
\begin{equation}\label{jetburst:sub:mu}
\mu=\frac{3f_D}{5}(\frac{4\pi}{3})^{10/9}\ (\frac{1}{\kappa})^{2/3}\
    (\frac{1}{\alpha c})^{1/9}\ \frac{1}{cT}\ (I{\cal E})^{5/9}\ (\frac{1}{\Sigma})^{10/9}\ R^{1/3}.
\end{equation}

As usual, the nearest ``on'' source pointing at us is likely to be
closer than $r_{\Sigma}$ and therefore detectable provided
$\mu>ln(2).$  This condition is here equivalent to
\begin{equation}\label{jetburst:sub:R1}
R\ >\ (\frac{5 ln 2}{3f_D})^3\ (\frac{3}{4\pi})^{10/3}\ \kappa^2\
  (\frac{1}{I\cal E})^{5/3}\ (\alpha c)^{1/3}\ (cT)^3\
  \Sigma^{10/3}.
\end{equation}

Substituting $0.117N$ for $I{\cal E}$, this becomes
\begin{equation}\label{jetburst:sub:R2}
\boxed{ R_{min} = \frac{0.47}{f_D^3} \ (\alpha c)^{1/3}\
  (\frac{\Sigma^2}{N})^{5/3}\ \kappa^2\ (cT)^3.}
\end{equation}

A fiducial calculation can be made again with the same parameter
values as before: $\kappa=0.014\ sr/Mpc,\ N=10,000,\ \alpha
c=1/(36\times 10^2),\ cT=1500\ Mpc,\ \Sigma=3,\ f_D=1/2$.  The minimum jet
burst fossil separation is needed for likely source detection is
$R_{min} = 1.3\ Mpc.$
Using $T\approx 4.5\times 10^9\ yrs$, 
the minimum mean time between bursts in a single galaxy is 
$<\Delta t>_{min}=0.01R^3T=1.0\times 10^8\ yrs$.  In the sub-GZK
regime, detecting the nearest source is unlikely unless the mean time
between bursts in a single galaxy is at least 100 million years.
Note, however, that this calculated minimum time is extremely
sensitive to the adopted value for $cT$, since it is proportional to
$R^3T$ and $R$ itself is proportional to $(cT)^3$.  A small change in
the adopted parameter value for $cT$ produces a dramatic change in the
minimum mean time interval.

\section{Luminosity function}

The basic model used in this paper assumes that cosmic ray sources are
all the same in the sense that there is a single luminosity $Q$
(particles emitted per unit time above the cosmic ray energy of
interest).  This is surely an overly simple idealization.  There
must be some distribution of luminosities $n(>Q)$, i.e. the spatial
density of cosmic ray sources with luminosity greater than $Q$.  

The luminosity function is constrained by the observed intensity
$I(>E)$ of cosmic rays.   The luminosity function must account for the
particle density $\frac{4\pi}{c}I$:
$$T\int \ Q\ \frac{dn}{dQ}\ dQ=\frac{4\pi}{c}I.$$
After imposing this normalization condition, a hypothetical luminosity
function $n(>Q)$ will generally not have a single parameter that could
be used in place of R.  The basic model in this paper has focused on
the mean separation R between identical sources.  There is no natural
generalization of the R-analysis for an arbitrary luminosity function.

Given any hypothetical luminosity function $n(>Q)$, however, one can
evaluate the probability $P$ that one or more cosmic ray sources is
detectable.  It is given by $P=1-e^{-\mu}$, where $\mu$ is the
expected number of detectable sources. Suppose a detector with
exposure $\cal E$ requires $K$ super-GZK events from some candidate
source (or signal-to-noise $\Sigma$ for sub-GZK events) to confirm a
detection.  At any distance $r$, there is a minimum $Q(r)$ that is
needed for that:
$$\frac{Q(r)\cal E}{4\pi r^2}\ >\ K\ \ \ \ or\ \ \ \ \frac{Q(r)}{4\pi
  r^{5/2}}\ \sqrt{\frac{\cal E}{\kappa I}}\ >\ \Sigma$$
for super-GZK or sub-GZK analysis, respectively.
The expected number of detectable sources is then 
$$\mu\ =\ \int_0^{\infty}\ 4\pi r^2\ n(>Q(r))\ dr,$$
which then gives the probability $P=1-e^{-\mu}$ that one or more
sources is detectable for the hypothetical luminosity function
$n(>Q)$. 

\section{Incomplete sky coverage}

If there are discrete sources of cosmic rays at all, they presumably
surround us.  The brightest nearby source could be in any part of the
sky.  It is important to achieve good exposure to the full celestial
sphere.  No cosmic ray observatory has so far been built with full-sky
exposure, although the Auger Observatory has been designed for that.

The analyses in this paper have used a simplifying approximation that
a single-site detector has the same good exposure to a fraction of the
sky denoted by $f_D$.  The southern site of the Auger
Observatory, by itself, has good exposure to approximately one quarter
of the sky at the most southern declinations, decreasing exposure over
half of the sky, and zero exposure to the northernmost quarter of the
sky.  Adding the complementary exposure from its northern site would
yield nearly uniform acceptance to cosmic rays from all parts of the
sky \cite{Sommers}.  

The formulas in this paper become more complicated for the real
situation in which exposure ${\cal E}(\vec{u})$ is not constant over
the celestial sphere but depends on direction $\vec{u}$.  (Here
$\vec{u}$ denotes a unit direction vector.)  In the case
of isotropic persistent sources, for example, the signal is 
$${\cal S}=\frac{IR^3}{cTr^2}{\cal E}(\vec{u}).$$
(See equation \ref{basic:signal}.)
For the super-GZK analysis, the expected signal is $K$ at radius $r_K$
given by
$$r_K^2=\frac{IR^3}{cTK}{\cal E}(\vec{u}).$$
The expected number of detectable sources out to distance $r_K$, which
now depends on direction $\vec{u}$, is
\begin{equation}\label{exposure:sup}
\mu=\int \int_0^{r_K(\vec{u})}\frac{1}{R^3}r^2 dr\
d\Omega=\frac{1}{3}(\frac{IR}{cTK})^{3/2}\ \int ({\cal E}({\vec{u}}))^{3/2}
d\Omega.
\end{equation}
The nearest source is still more likely to be detectable
than not provided $\mu>ln(2)$, which now becomes
$$R>(\ \frac{3ln2}{\int ({\cal E}({\vec{u}}))^{3/2} d\Omega}\ )^{2/3}\
\frac{cTK}{I}.$$  Similar modifications pertain to the sub-GZK
analysis.  Both the signal and the noise depend on direction $\vec{u}$,
and the ratio is 
$${\cal S}/{\cal N} = \sqrt{\frac{I{\cal E}(\vec{u})}{\kappa}}\ \frac{R^3}{cTr^{5/2}}.$$
The distance at which this has a prescribed value $\Sigma$ is given by
$$r_{\Sigma}^{5/2}=\sqrt{\frac{I}{\kappa}}\ \frac{R^3}{cT\Sigma}\
\sqrt{{\cal E}(\vec{u})}.$$  
The expected number of detectable sources is 
\begin{equation}\label{exposure:sub}
\mu=\int \int_0^{r_{\Sigma}(\vec{u})}\frac{1}{R^3}r^2 dr\
d\Omega=\frac{1}{3}(\frac{IR}{\kappa})^{3/5}\
(\frac{1}{cT\Sigma})^{6/5}\ \int ({\cal E}(\vec{u}))^{3/5} d\Omega.
\end{equation}
The condition $\mu>ln(2)$ becomes
$$R>(3\ ln2)^{5/3}\ \frac{\kappa}{I}\ (cT\Sigma)^2\ (\frac{1}{\int
  ({\cal E}(\vec{u}))^{3/5} d\Omega})^{5/3}.$$

A simple schematic approximation for the Auger South exposure 
is ${\cal E}={\cal E}_0 g(v)$ where $v\equiv \sin(declination)$,
${\cal E}_0$ is the rich exposure in the southern sky, and $g(v)$ is
the simple function
\[ g(v)\equiv \left\{ \begin{array}{ll}
1 & \mbox{if $-1<v<-1/2$} \\
1/2-v & \mbox{if $-1/2<v<1/2$} \\
0 & \mbox{if $v>1/2$}.  \end{array}
\right.  \]
(Note that $g(v)+g(-v)=1$, so identical sites in the north and south
provide uniform sky coverage ${\cal E}_0$ in this approximation.)

This schematic model of exposure for Auger South allows an approximate
analytic evaluation of $\int ({\cal E}(\vec{u}))^{\gamma}d{\Omega}$
for the various powers $\gamma$ which would arise in the different
models considered in this paper.  The resulting formulas are certainly
less transparent, however, than for the ideal detector with uniform
exposure.  

An effective sky coverage $f_D$ for Auger South can be defined by 
requiring that the expression for $\mu$ in equation \ref{exposure:sup}
be a scaled version of the expression in equation \ref{basic:sup:mu}, so
$$\frac{4\pi}{3}(\frac{I{\cal E}_0 R}{cTK})^{3/2} f_D\ \ =\ \ \frac{1}{3}(\frac{IR}{cTK})^{3/2}\ 2\pi\int ({\cal E}(v))^{3/2}dv.$$
This yields
$$f_D=\frac{2\pi}{4\pi}\int_{-1}^{+1}[g(v]^{3/2}dv\ =\ \frac{9}{20}.$$
For the sub-GZK case, the analogous condition on $f_D$ coming from
equations \ref{exposure:sub} and \ref{basic:sub:mu} leads to
$$f_D=\frac{2\pi}{4\pi}\int_{-1}^{+1}[g(v]^{3/5}dv\ =\ \frac{9}{16}.$$
These results suggest that 1/2 is a suitable estimate for the
effective sky coverage $f_D$ that pertains to a single-site
observatory.

\section{Summary and conclusions}

The likelihood of detecting a discrete source of ultra-high energy
cosmic rays depends on many variables with unknown values.  It has
here been assumed that cosmic rays are produced in sources that are
randomly distributed throughout the universe with some mean separation
$R$.  The analyses have focused on the question, ``What condition on
the mean separation ensures that the detection of the nearest source
has more than a 50\% chance of being detectable?''  The question is
simplified by assuming that all sources have the same luminosity $Q$.
The answer certainly depends on the amount of exposure $\cal E$ that
the detector has to the sources.  Since the number of arrival
directions in a data set increases in proportion to the detector's
celestial exposure, the answer can be regarded alternatively as
depending on the number $N$ of arrival directions collected above the
chosen energy cut.  It also depends on the cosmic ray accumulation
time $T$ for that energy cut.  Another variable that affects the
answer is the statistical significance required for detection -- the
required number of events $K$ in the signal-limited case (super-GZK)
or the required signal-to-noise ratio $\Sigma \equiv {\cal S}/{\cal
N}$ in the background-limited case (sub-GZK).  In addition, the answer
depends on whether the sources are permanent sources emitting
isotropically, permanent sources with beamed emission, isotropic
bursts, or jet bursts.  Magnetic fields also have an important impact
on the answer for the background-limited cases.

Properties of intergalactic magnetic fields are not well established.
For permanent sources, the relevant information about magnetic fields
is summarized by the coefficient $\kappa$ which governs how the
(Gaussian) solid angle of arrival directions grows with distance from
a source: $\omega=\kappa r$.  Appendix B examines how $\kappa$ can be
calculated from the mean field strength and coherence length in a
simple model.  

For transient (burst) sources, magnetic smearing of arrival times is
relevant as well as the smearing of arrival directions.  The time
smearing is encapsulated in the coefficient $\alpha$ by $\tau=\alpha
r$.  Appendix C shows how to estimate $\alpha$ from the magnetic field
properties in the simple model that is used to estimate $\kappa$.

The tables below collect the boxed formulas appearing in the text.
For each class of models, there are two formulas for the minimum mean
separation $R_{min}$ such that the probability of detection is greater
than 1/2.  One formula pertains to the super-GZK (signal-limited)
regime, and the other formula pertains to the sub-GZK
(background-limited) regime.

\begin{table}[htb]
\begin{center}
\caption{Formulas for $R_{min}$.  This is the minimum mean source
  separation for which one expects the nearest source to be
  detectable.   Formulas are tabulated for the signal-limited
  (super-GZK) case and the background-limited (sub-GZK) cases.  These
  formulas are highlighted by boxes in the text.  For a full-sky
  observatory with uniform exposure, $N$ should be half the number of
  arrival directions in the data set of events above the energy
  corresponding to the cosmic ray accumulation time $T$, and the
  fraction of sky exposed to the detector is $f_D=1$.  For a single
  site like Auger South, $N$ is the total number of events in the data
  set, and $f_D\approx 1/2$.  For non-uniform exposure, it would be
  better to replace $N$ by $I{\cal E}/0.117$ and use the exposure
  ${\cal E}$ that pertains to the target source celestial position and
  the chosen energy cut for which the cosmic ray intensity is $I$.}
\begin{tabular}{|c||c|c|c|}
\hline
Source Type & Super-GZK & Sub-GZK \\
\hline
\hline
Steady isotropic & $R_{min}=4.1\ \frac{K}{N}\ cT$ & $R_{min}=1.4\ (cT\ \Sigma)^2\ \frac{\kappa}{N}$ \\
\hline
Steady Jets & $R_{min}=4.1\ (\frac{\Omega_J}{4\pi})^{1/3}\ \frac{K}{N}\ cT$ & $R_{min}=1.6(\frac{\Sigma^2}{N})^{4/3}(cT)^{8/3}\kappa^{5/3}$ \\
\hline
Isotropic Bursts & $R_{min}=\frac{3.9}{f_D}(\alpha c)^{1/3}(\frac{K}{N})^{4/3} cT$ & $R_{min}=0.51(\frac{cT}{f_D})^{7/3} (\alpha c)^{1/3}
(\frac{\kappa \Sigma^2}{N})^{4/3}$ \\
\hline
Jet Bursts & $R_{min}=\frac{1.7}{f_D} (\alpha c \Omega_J)^{1/3}(\frac{K}{N})^{4/3}cT$ &  $R_{min} = \frac{0.47}{f_D^3} \ (\alpha c)^{1/3}\
  (\frac{\Sigma^2}{N})^{5/3}\ \kappa^2\ (cT)^3$ \\
\hline
\end{tabular}
\end{center}
\end{table}

A statistical detection of multiple sources can be expected prior to
the detection of any one discrete source.  Study of the two-point
correlation function (or, equivalently the angular power spectrum) for
ultra-high-energy cosmic ray arrival directions can exhibit evidence
for many poor clusters of arrival directions even if there is not any
one rich cluster of arrival directions that is individually detectable
\cite{Olinto}.  Evidence for an autocorrelation in the AGASA data was
published \cite{AGASA}.  Moreover, a correlation of arrival directions
with a catalog of candidate sources can supply evidence of discrete
sources even if there is no statistical evidence for clustering of the
cosmic rays themselves.  Exploratory searches have also produced
evidence for correlations of that type \cite{Tinyakov}.  These
specific claims will be thoroughly tested using larger data sets obtained
with new observatories.  The failure to detect any individual discrete
source so far suggests that there is no really bright cosmic ray
source in the sky.  It is therefore likely that sources will show up
collectively in one of these ways before any individual source becomes
obvious.

Whether or not the brightest cosmic ray source will soon become
detectable depends on many unknown parameter values.  The issue must
be decided observationally.  Fiducial estimates show that it is a
close call.  The answer can go either way, depending on assumptions
about the unknown parameter values and the nature of the sources
(e.g. bursts, jets).  Present detector exposures are already obtaining
important constraints on the unknown astrophysical parameters.  The
viable parameter space will shrink rapidly as the exposure increases.
The search will be especially rewarding, however, if it leads to the
study of one or more sources and the intervening magnetic fields.

\section{Acknowledgments}
I am grateful to Mike Roberts for many discussions on these issues and to 
Etienne Parizot for useful comments on an early draft.
The work has been supported in part by NSF grant PHY-0555317.

\section{Appendix A: Minimum and maximum $R$} 

The analysis in the text focuses on the mean separation $R$ between
identical sources (density of sources $n=1/R^3$).  The nearest source
is unlikely to be detectable if $R$ is too small, because then each
source is too weak.  For super-GZK analysis, there is an upper
limit $R_{max}$ as well as lower limit $R_{min}$ stemming from the
fact that the nearest source is unlikely to be much closer to us than
the mean separation $R$, so $R$ cannot be much greater than $cT$,
where $cT$ is the effective survival distance for those super-GZK
particles. 

In a volume of radius $r$ around us, the number of detectable sources
is 
\begin{equation}\label{appendixA:1}
\mu=\frac{4\pi f_D}{3}r^3/R^3.
\end{equation}
As explained in section 3, the nearest source is likely to be
detectable provided $\mu>ln 2$.  The critical value $\mu=ln 2$ gives
$R_{min}$ and $R_{max}$.  Equation \ref{appendixA:1} shows that these
are simply proportional to the corresponding volume radii:
$$ R_{min}=\lambda r_{min}\ \mbox{ and } R_{max}=\lambda r_{max},$$
where the proportionality constant is 
$$\lambda=(\frac{4\pi f_D}{3 ln 2})^{1/3}.$$

Equation \ref{basic:luminosity} gives the luminosity $Q=\frac{4\pi I
  R^3}{cT}$ per source with mean separation $R$ which collectively
  account for the cosmic ray density $4\pi I/c$.  The signal $\cal S$
  seen at distance $r$ after exposure $\cal E$ in the basic
  (isotropic) model is 
$${\cal S}=\frac{I{\cal E}R^3}{cTr^2}e^{-r/cT},$$
where the GZK attenuation factor $e^{-r/cT}$ is here included.  If $K$
events are required for a source detection, the critical case is
obtained by setting ${\cal S}=K$.  Measure distances relative to $cT$,
so $r=\xi cT$ and $R=\lambda \xi cT$ when the distance $r$ is
$r_K=r_{min}$ or $r_K=r_{max}$.  Then the equation above for ${\cal
  S}$ reduces to 
\begin{equation}\label{appendixA:2}
\xi e^{-\xi}=\frac{K}{\lambda^3 I {\cal E}}.
\end{equation}
 A solution for
$\xi$ requires $\frac{K}{\lambda^3 I {\cal E}}\leq \frac{1}{e}$, and
there are two solutions except if the equality holds.  One solution
has $\xi < 1$ (i.e. $r_K<cT$), and the other has $\xi > 1$
(i.e. $r_K>cT$).  For cases in which $\frac{K}{\lambda^3 I {\cal E}}
\ll 1/e$, then $e^{-\xi}\equiv e^{-r_K/cT} \cong 1$.  For $r<r_K$, the
attenuation factor $e^{-r_K/cT}$ is then very close to 1, and the
simpler analysis following equation \ref{basic:signal} is fully
justified.

The approximation $I{\cal E}=0.117N$ introduced in section 3 can be
used here.  Adopting $f_D=1/2$ as in the fiducial calculations, one
gets
$$\frac{K}{\lambda^3 I{\cal E}}<1/e \Leftrightarrow K<0.13N.$$
Therefore, provided the data set has more than 8 times the number of
arrival directions $K$ needed for detection of the discrete source,
the analysis in the text (omitting the $e^{-r/cT}$ attenuation factor)
is adequate.

The upper limit $R_{max}=\lambda \xi cT$ is given by the other
solution of equation \ref{appendixA:2} for which $\xi > 1$.

\section{Appendix B: Blurring by magnetic fields}

The Larmor radius $\rho$ characterizes the bending of charged
particle trajectories by magnetic fields.  For relativistic particles,
it is given by

\begin{center}
\[ \rho_{cm}=\frac{E_{eV}}{300\ Z\ B_G} \Rightarrow  \left\{ \begin{array}{ll}
        \rho_{Mpc}\ \dot{=}\ \frac{E_{EeV}}{Z B_{nG}}  &
        \mbox{for extragalactic applications};\\
	& \\
        \rho_{kpc}\ \dot{=}\ \frac{E_{EeV}}{Z B_{\mu G}} & \mbox{for galactic
        applications.}\end{array} \right. \]  
\end{center}
The approximate equations on the right follow from the exact equation
on the left using the relations: $1\ EeV\equiv 10^{18}\ eV,\ 1\
nG\equiv 10^{-9}\ G,\ 1\ Mpc\dot{=}3.1\times 10^{24}\ cm,\ 1\ \mu
G\equiv 10^{-6}\ G,\ 1\ kpc\dot{=}3.1\times 10^{21}\ cm.$ The
particle's electric charge $Z$ is in units of 1 proton charge.  In
traveling distance D through a perpendicular B-field, a particle's
trajectory is bent by the angle $\theta=D/\rho$ in radians.   

The Galaxy has a regular magnetic field which tends to follow the
spiral arms, and also superposed irregular fields which change
direction over short distances along any path.  Particles arriving
from a distant point source will be systematically deflected by the
regular field.  An estimate is that they will encounter, on average, a
perpendicular magnetic field of about $2\ \mu G$ acting over a path of
roughly 1 $kpc$.  Trajectories from a single source would then be
systematically bent through the angle
$$\theta=\frac{1\ kpc}{\rho}=\frac{(1\ kpc) Z (2\ \mu G)}{E_{EeV}}\ radians.$$
For protons ($Z=1$) this is about $12^{\circ}$ at 10 $EeV$ and $1.2^{\circ}$
at 100 EeV.

A uniform field acting along the entire path from source to detector
would cause the arrival direction to differ from the source direction
by only $\theta/2$.  This is because, in that case, the particle does
not start from the source in our direction; its initial direction also
differs from our line of sight by half of the trajectory bending angle
($\theta/2$).  For a distant extragalactic source, however, the
regular field changes the arrival direction by the full angle $\theta$
from the particle's direction of entry into the galaxy (which is the
direction from detector to source if the source is very distant and
the particle travels on a straight line until reaching the Galaxy).

All protons of one energy are deflected the same amount by the regular
magnetic field.  With arrival directions of two or more protons of
measured energies from the same source, one can determine the product
$BD$ and the source direction.  Here $BD$ is the transverse
magnetic field integrated over its range along the incoming trajectory.
The regular field produces an arc of
arrival directions on the sky ending at the source direction
($E=\infty$) with $E$-dependent angular offsets of $\theta=BD/E$.  

Because of the steep energy spectrum, it can be expected that roughly
3/4 of the arriving particles will have energy less than twice the
analysis threshold energy.  Using the above estimate of deflection by
the regular galactic magnetic field, one would expect that above 10
EeV, for example, the regular field should spread 75\% of the arrival
directions along an arc of roughly 6 degrees (plus or minus 3
degrees), with the center of that arc displaced from the source
direction by approximately 9 degrees.  Above 100 EeV, the arc would be
roughly 0.6 degree (plus or minus 0.3 degree) with its center
displaced from the source by approximately 0.9 degree.  The input
values for these estimates (2 $\mu G$ for the transverse regular field
strength and 1 $kpc$ for the effective path length) are crude
estimates, and actual values depend critically on the direction to any
given source.  This rough estimated does indicate, however, that
clusters of arrival directions should not be destroyed by the Galaxy's
regular magnetic field for most source directions.

Irregular magnetic fields are the other concern in charged particle
astronomy.  These are fields that do not have a consistent direction
over any particle's trajectory.  The particle's direction is
continuously being deflected by small magnetic bends that cause it to
meander in a random-walk manner.  The result is formally the same as
multiple coulomb scattering of energetic charged particles in
matter. A uni-directional initial beam becomes a Gaussian distribution
of particle directions centered on the undeflected direction.  The
width of the Gaussian distribution ($\sigma$) increases in proportion
to the square root of the path length $r$.  Appendix C shows that the
effective solid angle is ($\omega=4\pi \sigma^2$), which increases in
proportion to the path length.  The proportionality constant $\kappa$
is defined by $\omega=\kappa r$.

A simple model of magnetic blurring by irregular fields is that a
particle encounters a different field orientation in each segment of
length $L$ along its path.  The field has a mean strength $B$ with a
random orientation which is constant over each segment.  Consider
deflection in any plane containing the original direction.  The
B-component perpendicular to that plane is expected to be
$B/\sqrt{3}$.  Over a segment length $L$ the deflection in the plane
is
$$\theta_s=\frac{LZB}{\sqrt{3}E}.$$  This is the random
walk step size.  After $n:=r/L$ steps, the directions in that plane
are distributed with a Gaussian of width $$\sigma=\sqrt{n}
\theta_s=\sqrt{\frac{r}{L}} \frac{LZB}{\sqrt{3}E}=\sqrt{\frac{rL}{3}}\
\frac{ZB}{E}.$$  

This is the distribution of directions relative to an original beam
direction.  What is relevant is the direction of a detected particle
relative to the direction from the detector to the source.  That
direction to the source does not correspond to the original particle
direction because the particle will have been displaced laterally from
the beam.  The lateral displacement is correlated with the offset of
the arrival direction.  As in multiple Coulomb scattering, the arrival
directions are distributed around the direction to the beam origin
with a Gaussian $\sigma$ which is smaller by $1/\sqrt{3}$.  This is a
straightforward consequence of the statistical correlation of angular
offset with spatial offset in the random walk process.  The direction
back to the source is different for particles that have a net
deflection to the left than for for those that have a net deflection
to the right, and the final direction is statistically correlated with
the net deflection.

Using $\kappa=\omega/r$ together with $\omega=4\pi \sigma^2$ and
$\sigma=\sqrt{\frac{r}{L}} \frac{\theta_s}{\sqrt{3}}$ yields a formula
for $\kappa$:
$$\kappa = \frac{4\pi}{9}\ \frac{LZ^2B^2}{E^2}.$$

This expression for $\kappa$ is used in the text for evaluating the
magnetic blurring due to random intergalactic magnetic fields between
a distant source and the Earth.  There is also a contribution by
irregular magnetic fields within the Galaxy, and this expression for
$\kappa$ can be used to estimate its magnitude.  Suppose $B=3\ \mu G$
for the randomly-oriented field strength, that $L=0.1\ kpc$, and
the travel path through those irregular galactic fields has length $r=1\
kpc$.  Then the Gaussian spread $\sigma$ is given (for $Z=1$) in
radians by
$$\sigma=\sqrt{\omega/4\pi}=\sqrt{\frac{LZ^2B^2r}{9E^2}}= \frac{0.32}{E}.$$
Setting $E=10\ EeV$ gives $\sigma=1.8^{\circ}$.

\section{Appendix C: Time spreading by magnetic fields}

The simple model of irregular fields used in Appendix B provides an
estimate for the coefficient $\alpha$ that governs the time spread $\tau$
for a source at distance $r$ by $\tau=\alpha r$.  The estimate for $\tau$ is
based on the expected difference in transit time for a charged
particle of energy $E$ compared to an undeflected neutral
speed-of-light particle.  There is a time difference in each path segment
of length $L=\rho \theta$ because the curved trajectory is longer than the
straight line distance between the endpoints.  (Here $\rho$ is the Larmor
radius based on the perpendicular magnetic field, and theta is the
trajectory bending angle while traveling distance $L$.)  The time
difference is
$$\Delta t = \frac{1}{c}(L-2 \rho sin(\theta/2)) = \frac{\rho}{c}(\theta-2
sin(\theta/2))\approx \frac{\rho}{c}\frac{\theta^3}{24} =
\frac{L^3}{24c\rho^2}.$$
The total time difference is then
$$\tau = \frac{r}{L} \Delta t = \frac{L^2}{24R^2}\ \frac{r}{c}.$$
Using $R=E/ZB_{\perp}$ and $< B_{\perp}^2 > = \frac{2}{3}B^2$,
this gives the coefficient $\alpha \equiv \tau/r$:
$$\alpha = \frac{L^2Z^2B^2}{36 c E^2}.$$  Here $B$ is the 
magnetic field strength whose direction is randomly oriented along
each trajectory segment of length $L$. 

This estimate ignores a second-order contribution to $\tau$ due to the segments
themselves meandering about the straight line from the source to the
arrival point.  

\section{Appendix D: Signal and noise with Gaussian-distributed arrival directions}

Random intergalactic magnetic fields are expected to produce a
Gaussian distribution of arrival directions at the detector, centered
on the source direction.  Suppose an observed source has arrival
directions distributed about a central direction with a Gaussian of
width $\sigma$.  This means in any one dimension the probability
distribution for offset $\theta_x$ is $P(\theta_x) =
\frac{1}{\sqrt{2\pi}\sigma } exp(-\theta_x^2/2\sigma^2),$ and the
2-dimensional (space angle) offset
$\theta=\sqrt{\theta_x^2+\theta_y^2}$ has probability $P(\theta) =
\frac{1}{2\pi \sigma^2}exp(-\theta^2/2\sigma^2).$ When testing a
discrete source with an apparent Gaussian distribution of arrival
directions, it is sensible to give more weight to arrival directions
that are near the center of the distribution and little weight to
arrival directions that are far from it.  The appropriate weighting
function is $$w=4\pi \sigma^2 P(\theta) = 2exp(-\theta^2/2\sigma^2).$$
This choice enjoys three important properties: \\

\noindent(1) The {\it shape} of the Gaussian function of width
$\sigma$ maximizes the signal among all possible weighting functions
of the same integral normalization.\\

\noindent(2) With the $4\pi \sigma^2$ integral normalization, the
weighted integral of any uniform background has expected fluctuations
given by the square root of the background itself.  This expected
background is $4\pi \sigma^2 \times (background\ density)$, and the RMS
fluctuation in that weighted background integral is its square root.
\\

\noindent(3) With the $4\pi \sigma^2$ integral normalization, a
density $N_0 P(\theta)$ of smeared-out arrival directions due to $N_0$
cosmic rays from a discrete source results in a weighted integral
equal to $N_0$, i.e. the actual number of smeared-out directions.\\

Defining signal and background as weighted integrals with this
weighting function, their difference is the expected number of events
producing the signal.  Moreover, the usual $S/N$ statistical
significance pertains with the noise $\cal N$ being simply the square root
of the background, as in analyses without a weighting function.
One can regard $2\sigma$ as an effective radius, giving $4\pi
\sigma^2$ as an effective collecting area.  

A more careful analysis should use Fisher distributions \cite{Fisher}
rather than Gaussians for celestial analyses, especially for the broad
distributions that are expected in sub-GZK analyses.

\end{document}